# Observation of Nonlinear Spin-Charge Conversion in the Thin Film of Nominally Centrosymmetric Dirac Semimetal SrIrO$_3$ at Room Temperature

Y. Kozuka,[1]* S. Isogami,[1] K. Masuda,[1] Y. Miura,[1] Saikat Das,[1] J. Fujioka,[2] T. Ohkubo,[1] S. Kasai[1,3]

[1]*Research Center for Magnetic and Spintronic Materials, National Institute for Materials Science (NIMS), 1-2-1 Sengen, Tsukuba 305-0047, Japan*

[2]*Faculty of Material Science, University of Tsukuba, Tsukuba, Ibaraki 305-8571, Japan*

[3]*Japan Science and Technology Agency, PRESTO, Kawaguchi, Saitama 332-0012, Japan*

Abstract

Spin-charge conversion via spin-orbit interaction is one of the core concepts in the current spintronics research. The efficiency of the interconversion between charge and spin current is estimated based on Berry curvature of Bloch wavefunction in the linear-response regime. Beyond the linear regime, nonlinear spin-charge conversion in the higher-order electric field terms has recently been demonstrated in noncentrosymmetric materials with nontrivial spin texture in the momentum space. Here we report the observation of the nonlinear charge-spin conversion in a nominally centrosymmetric oxide material, SrIrO$_3$, by breaking inversion symmetry at the interface. A large second-order magnetoelectric coefficient is observed at room temperature because of the antisymmetric spin-orbit interaction at the interface of Dirac semimetallic bands, which is subject to the symmetry constraint of the substrates. Our study suggests that nonlinear spin-charge conversion can be induced in many materials with strong spin-orbit interaction at the interface by breaking the local inversion symmetry to give rise to spin splitting in otherwise spin degenerate systems.

*corresponding author e-mail: KOZUKA.Yusuke@nims.go.jp

Interconversion of electron spins and charges is one of the central techniques in the current spintronic research. Electrons accelerated by the electric field in a nonmagnetic metal are deflected through spin-orbit interaction, generating spin current transverse to the charge current [1,2]. The spin current can be utilized to flip the magnetic moment in the adjacent ferromagnets and applied for electrical magnetization switching with low energy consumption in spintronic devices [3-5]. The efficiency of charge-to-spin conversion has been estimated by calculating spin Hall conductivity based on the Berry phase approach [1,6] or equivalently Kubo formula in the linear-response regime [7]. In this approach, a band degeneracy point acts as a source of Berry curvature, analogous to the magnetic field in the momentum space, and gives a large contribution to spin or anomalous Hall effects [2,8].

Recently, higher-order spin-charge conversion beyond the linear-response theory has been recognized in noncentrosymmetric materials both theoretically and experimentally, where the electric field is coupled with Berry curvature dipole, leading to the observation of second-order nonreciprocal Hall effect [9-15]. Nonlinear spin-charge conversion is also found to emerge under the in-plane magnetic field for the topological surface states of a topological insulator [16]. This nonlinear planar Hall effect originates from the transverse shift of spin-momentum locked topological surface states under the in-plane magnetic field in the presence of a nontrivial $k$-cubic warping effect. Conversely, it is suggested that nonlinear planar Hall (and nonlinear magnetoresistance as well) can be utilized to probe the nontrivial spin texture in the momentum space in several materials without relying on spin- and angle-resolved photoemission spectroscopy [16,17].

Here we report a large nonlinear spin-charge conversion detected by a harmonic measurement of the planar Hall effect in nominally centrosymmetric SrIrO$_3$ thin films at room temperature. The orthorhombic phase of SrIrO$_3$ is known to be a Dirac semimetal with comparable spin-orbit interaction and electron correlation (~0.5 eV) with a Dirac nodal ring about 50 meV below the Fermi energy formed from two Dirac bands [18-21]. Owing to the characteristic band structure, this material is known to show a large spin Hall effect and to generate strong spin-orbit torque to the adjacent ferromagnetic layer [22-26]. The crystal structure of SrIrO$_3$ is distorted perovskite (GdFeO$_3$-type) in the centrosymmetric *Pbnm* space group with lattice constants of $a$ = 5.60 Å, $b$ = 5.58 Å, $c$ = 7.75 Å (corresponding to a pseudo-cubic lattice constant of $a_{\text{pseudo}} = \sqrt{(2a^2 + 2b^2 + c^2)/12} = 3.93$ Å [Fig. 1(a)] [27]. In the presence of inversion symmetry, the band structure of SrIrO$_3$ maintains spin degeneracy, and nonlinear spin-charge conversion is not expected. However, the spatial inversion symmetry of thin films is inherently broken at the surface and the interface, which causes spin textures in the momentum space due to antisymmetric spin-orbit interaction. Comparing the nonlinear planar Hall effect of SrIrO$_3$ thin films grown on different substrates, we conclude that the observations are interface/surface-driven phenomena irrespective of the degree of the strain with minor modification of spin texture by the symmetry constraint of the substrate.

The experimental details are explained in the supplemental material. Briefly, the SrIrO$_3$ thin films are grown at 650 °C under 100 mTorr oxygen partial pressure by pulsed laser deposition. We have employed LSAT (001) (cubic, $a$ = 3.87 Å), GdScO$_3$ (110) (orthorhombic, $a_{pseudo}$ = 3.96 Å), and NdGaO$_3$ (110) (orthorhombic, $a_{pseudo}$ = 3.86 Å) substrates to compare effects of epitaxial strain originating from lattice mismatch and crystal symmetry. The thicknesses are 25 nm, 27 nm, and 37 nm for thin films grown on LSAT (001), GdScO$_3$ (110), and NdGaO$_3$ (110), respectively. The crystal structures of SrIrO$_3$ and LSAT are depicted in Fig. 1(a) [28]. The second-order term of the planar Hall effect is characterized by an out-of-phase second-harmonic component using lock-in amplifiers with an alternating current frequency of 33 Hz [17]. The first-principles density-functional calculations including the spin-orbit interaction are carried out with the aid of the Vienna ab initio simulation program (VASP) [29-31].

Figure 1(b) shows a $\theta$-$2\theta$ scan of X-ray diffraction for SrIrO$_3$ thin films grown on LSAT (001), GdScO$_3$ (110), and NdGaO$_3$ (110) substrates, indicating epitaxial growth with out-of-plane lattice constants of 3.99 Å for LSAT, 3.89 Å for GdScO$_3$, and 3.97 Å for NdGaO$_3$, which deviate from the bulk value to compensate in-plane compressive (for LSAT and NdGaO$_3$) and tensile (for GdScO$_3$) strains (reciprocal space mapping shown in Fig. S1). The crystal structure is additionally characterized by high-angle annular dark-field transmission electron microscope (HAADF-STEM) images as shown in Fig. S2. While all films show clear epitaxial growth on the substrates, dislocations are partly observed for LSAT and NdGaO$_3$ substrates probably due to large lattice mismatch and presence of twins. The resistivity shown in Fig. 1(c) is only weakly temperature dependent as typical semimetallic behavior of SrIrO$_3$, consistent with other PLD-grown films in previous reports [32-34].

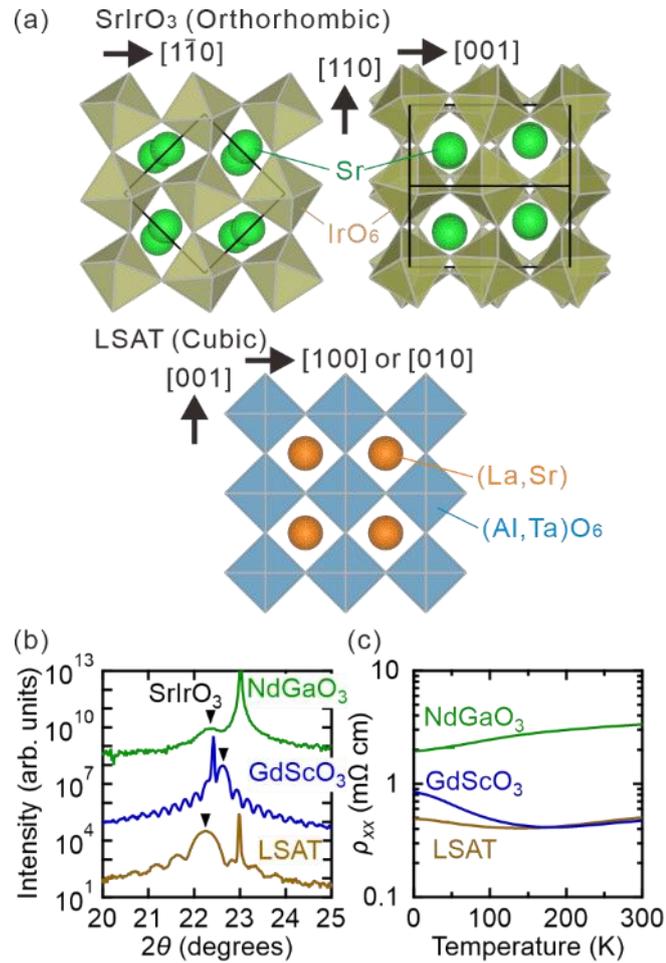

FIG. 1 . (a) Schematic diagram of crystal structures for (top) distorted orthorhombic perovskite (SrIrO$_3$) and (bottom) cubic perovskite (LSAT) materials. GdScO$_3$ and NdGaO$_3$ also have the orthorhombic structure but with different lattice parameters. Black lines in the orthorhombic structure indicate a unit cell. (b) $\theta$-$2\theta$ scan of X-ray diffraction and (c) temperature dependence

of resistivity for SrIrO$_3$ thin films grown on LSAT (001), GdScO$_3$ (110), and NdGaO$_3$ (110). The triangles in (b) show peaks of SrIrO$_3$ thin films.

To characterize the nonlinear planar Hall effect, we measure angular dependence ($\varphi$) of second-harmonic Hall voltage on in-plane magnetic field direction with respect to the current direction using a Hall bar structure as shown in Fig. 2(a). Figure 2(b) shows a typical second-harmonic planar Hall measurement as a function of $\varphi$ for a film grown on an LSAT substrate at 300 K under a constant magnetic field (*H*) of 10 T. The second-harmonic resistance is defined as $R_{yx}^{2\omega} = V_{yx}^{2\omega}/I$, where $V_{xy}^{2\omega}$ is the out-of-phase second-harmonic voltage and *I* is the externally applied alternating current. The data in Fig. 2(b) clearly shows $\cos\varphi$ dependence. So far, a similar cosine dependence of the second-harmonic planar Hall effect has been ascribed to nontrivially spin-momentum locked band structures in noncentrosymmetric materials [17] although, at first sight, this mechanism is not compatible with centrosymmetric SrIrO$_3$.

In order to obtain an insight into the origin, the magnitude of the second-harmonic planar Hall signal ($\Delta R_{yx}^{2\omega}$) is extracted by fitting with $R_{yx}^{2\omega} = \Delta R_{yx}^{2\omega} \cos\varphi$ (after subtracting a constant background) [17]. First, as shown in Fig, S3(a) (supplemental material), we confirm that $\Delta R_{yx}^{2\omega}$ is proportional to alternating current [Fig. 2(c)], which indicates the second-harmonic voltage $V_{yx}^{2\omega}$ certainly captures a second-order signal proportional to $I^2$.

Subsequently, we investigate magnetic field dependence [Fig. S3(b) of the supplemental material] and find that $\Delta R_{yx}^{2\omega}$ is proportional to $H$ [Fig. 2(d)]. These results are consistent with the nonlinear planar Hall effect, which is proportional to both the electric field ($E_x$) and magnetic field ($H$) as previously observed in $Bi_2Se_3$ [16,17]. We have also confirmed the second-harmonic component of anisotropic magnetoresistance ($R_{xx}^{2\omega}$) as shown in Fig. S4. The ratio between transverse and longitudinal nonlinear magnetoelectric effect $\rho_{yx}^{2\omega}/\rho_{xx}^{2\omega} = (R_{yx}^{2\omega}/R_{xx}^{2\omega})(L/W)$ (L: channel length = 25 μm, W: channel width = 10 μm) is estimated as 0.50, and the contribution from the Nernst effect, proportional to $\boldsymbol{H} \times \nabla T$, can be ignored.

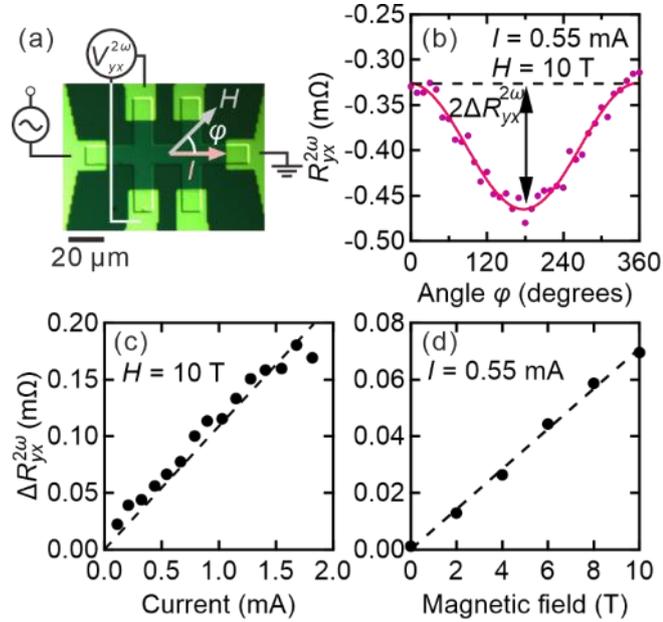

FIG. 2. (a) Microscope image of a Hall bar with Ti/Au electrodes together with the measurement geometry. The angle between the current (*I*) direction and in-plane magnetic field

($H$) is defined as $\varphi$. (b) Second-harmonic planar Hall resistance $R_{yx}^{2\omega} = V_{yx}^{2\omega}/I$ as a function of $\varphi$ and fitting with $\Delta R_{yx}^{2\omega} \cos \varphi$. (c) Current and (d) magnetic field dependences of $\Delta R_{yx}^{2\omega}$.

The bilinear magnetoelectric effect is known to capture complex spin-momentum locked band structures including warping effect or electron-hole asymmetry. The band structure of bulk SrIrO$_3$ is composed of two Dirac bands around the U point $(\pi, 0, \pi)$ and massive hole-like bands around $(0,0,0)$ and $(\pi, \pi, \pi)$ in the Brillouin zone (in the representation of the orthorhombic phase), both of which are spin degenerate due to the centrosymmetric crystal structure [19,20]. In the case of thin film, however, antisymmetric spin-orbit interaction induces momentum-dependent spin splitting at the interface, leading to complex spin texture in the two Dirac bands. Indeed, the nonlinear planar Hall effect was also observed at the LaAlO$_3$/SrTiO$_3$ interface [17] which is known to possess complex spin texture originating from the anti-crossing of three $t_{2g}$ bands [35,36].

To test this consideration, we compare the second-harmonic planar Hall signal for SrIrO$_3$ thin films grown on NdGaO$_3$ (110) and GdScO$_3$ (110) substrates as well as a film grown on LSAT (001) substrate [the raw data of the nonlinear planar Hall effect for GdScO$_3$ with $I \parallel$ [001] are shown in Figs. S3(c)–S3(f)]. Figure 3 shows the $\varphi$ dependence of $\rho_{yx}^{2\omega}$ normalized by $E_x$ and $H$, where $\rho_{yx}^{2\omega} = R_{yx}^{2\omega} d$ is the second-harmonic resistivity in the three-dimensional unit ($d$: thickness of the film). The magnitudes of the second-harmonic signal remarkably vary with different substrates. Furthermore, in the case of orthorhombic NdGaO$_3$ and GdScO$_3$ substrates, we find a significantly large anisotropy between two current directions along [001] and [1$\bar{1}$0]. For fair comparison for films with different resistivity, we use a coefficient of the bilinear magnetoelectric effect $\chi_{yxx} = \Delta\rho_{yx}^{2\omega}/E_x H$, where $\Delta\rho_{yx}^{2\omega} = \Delta R_{yx}^{2\omega} d$, which is shown in Table I [17]. We find that the room temperature $\chi_{yxx}$ value for the sample grown on LSAT is almost comparable to the previously reported value for a Bi$_2$Se$_3$ thin film at 5 K ($\chi_{yxx} \approx$ 0.02 mΩ μm$^2$/VT ), which rapidly decreases with increasing temperature toward room temperature. Remarkably, $\chi_{yxx}$ is significantly large for orthorhombic GdScO$_3$ and NdGaO$_3$ with the current along [001] but negligibly small along [1$\bar{1}$0]. Furthermore, the sign of $\chi_{yxx}$ differs between two current directions in the case of the GdScO$_3$ (110) substrate. These results indicate that the nonlinear planar Hall signal is robust and immune to strain and degree of disorder. The magnitude is sensitive to symmetry constraints from the substrates with minor modulation by the degree of strain, suggestive of an intrinsic effect reflecting the symmetry of the spin texture on the Fermi surfaces. For comparison, we also study a SrRuO$_3$ film grown on LSAT (001) substrate, which does not exhibit nonlinear planar Hall effect at room temperature as shown in Fig. S6. This signals the significance of the strong spin-orbit interaction and the characteristic band structure of SrIrO$_3$.

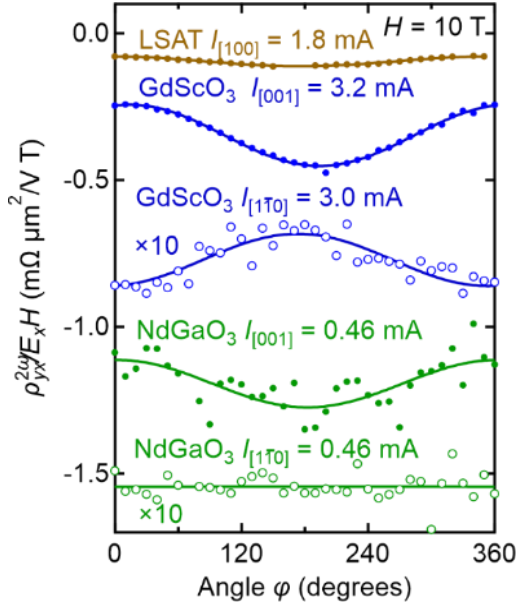

FIG. 3. Second-harmonic planar Hall resistivity $\rho_{yx}^{2\omega}$ ($= R_{yx}^{2\omega}d$, $d$: thickness) normalized by electric field along the current direction ($E_x$) and in-plane magnetic field ($H$) as a function of $\varphi$ for SrIrO$_3$ films grown on three kinds of substrates. For orthorhombic substrates (GdScO$_3$ and NdScO$_3$), measurements with two current directions are shown. The data for GdScO$_3$ and NdGaO$_3$ with the current along the [1$\bar{1}$0] directions are magnified by a factor of ten because of the small signal.

TABLE I. Comparison of resistivity and a coefficient of the nonlinear planar Hall effect $\chi_{yxx} = \Delta\rho_{yx}^{2\omega}/E_xH$ for SrIrO$_3$ films grown on various substrates. The data for two distinct current directions are shown for orthorhombic GdScO$_3$ and NdGaO$_3$ substrates.

| Substrate | Current direction | $\rho_{xx}$ (mΩ cm) | $\chi_{yxx}$ (mΩ μm²/VT) |
|---|---|---|---|
| LSAT (001) | [100] | 0.50 | 0.017±0.005 |
| GdScO$_3$ (110) | [001] | 0.47 | 0.10±0.004 |
|  | [1$\bar{1}$0] | 0.52 | -0.009±0.001 |
| NdGaO$_3$ (110) | [001] | 3.3 | 0.093±0.01 |
|  | [1$\bar{1}$0] | 3.3 | 0.0±0.0001 |

In the previous studies, band structure modification in SrIrO$_3$ thin films has been discussed in terms of epitaxial strain and geometric confinement, which causes an enhancement of spin Hall effect in the linear-response regime owing to preferable redistribution of Berry curvature [22,25]. In the present study, the large anisotropy of the nonlinear planar Hall effect reflects the spin-momentum locked Fermi surface composed of two correlated Dirac bands. Since the Fermi surface is deformed to the direction transverse to the external current in the planar Hall geometry with $I \parallel H$ [17], the large $\chi_{yxx}$ with $I \parallel [001]$ indicates that spin-momentum locking is stronger along [1$\bar{1}$0] than along [001]. Since [1$\bar{1}$0] is not equivalent to [$\bar{1}$10] in the orthorhombic structure, this anisotropic spin-momentum locking might be because of the lower in-plane symmetry along [1$\bar{1}$0] than along [001], which is crystallographically equivalent to [00$\bar{1}$] [Fig. 1(a)]. On the contrary, such anisotropy is not present in the case of the cubic LSAT substrate but an orthorhombic twin structure is known to appear based on the detailed structural analysis of SrIrO$_3$ thin films reported in [37]. This twin structure may average out the anisotropic nonlinear planar Hall signal in the case of the LSAT substrate.

To confirm this hypothesis, we have performed first-principles calculations of electronic band structure using an unstrained $SrIrO_3$ slab model [Fig. 4(a)] (bulk and strained case are shown in the supplemental material for comparison). The band structure is shown in Figs. 4(b) and 4(c). As shown in Fig. 4(c) (magnification around $\bar{U}$), inner Ir layers contribute to bands somewhat above the Fermi level, which resembles bulk Dirac bands, whereas the surface top and bottom Ir layers forming Fermi surfaces are expected to be subject to symmetry breaking. To obtain an insight into the spin-resolved band structure around the $\bar{U}$ point [Fig. S7(d)], Fermi surfaces are drawn together with local spins at each $k$ point for the unstrained slab model, projected to Ir atoms in the top layer as shown in Fig. 4(d). The spin texture is found as the surface states originating from the top Ir layer of the slab but summing the spins over the Fermi surfaces exactly compensates the spins, reflecting nonmagnetic nature. Moreover, the direction of the spins is inverted between the orbitals from the top and bottom Ir layers [Fig. S8(d)] at each $k$ point. These features are reminiscent of spin structures in the presence of antisymmetric spin-orbit interaction. Compared with conventional semiconductors, spin texture in $SrIrO_3$ is much more complex because spin-orbit interaction plays an essential role in the ground state of the electronic states as represented by the existence of Dirac points.

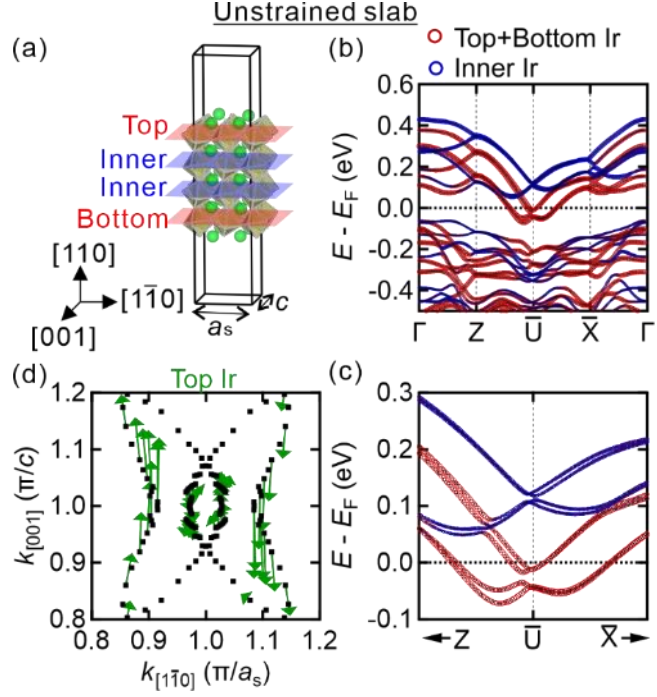

FIG. 4 . (a) Crystal structure and (b) band structure of the unstrained slab model. (c) The band structure is magnified around $\bar{U}$. In the panes (b) and (c), contributions from the surface (top and bottom) Ir layers (red) and from inner Ir layers (blue) are separately drawn by color. (d) Fermi surfaces and spin textures around the $\bar{U}$ point, where the contribution from top Ir layers is projected. Comparisons with bulk and strained slab are shown in the supplemental material in detail.

The obtained spin texture can be compared with the nonlinear planar Hall effect in Fig. 3. In the case of the unstrained slab model, the larger spins are found along $[1\bar{1}0]$, while spins are nearly canceled along $[001]$ direction as shown in Fig. 4(d). Since the nonlinear planar Hall effect detects the spin texture orthogonal to the current direction, the larger signal with $I \parallel [001]$ is consistent with the calculated spin texture. When the SrIrO$_3$ thin film is strained to GdScO$_3$ (110) substrate [38], the Fermi surfaces are enlarged because of the stretched Ir-O-Ir bonds as shown in Figs. S8(e)–S8(h), but the qualitative features are similar to the unstrained case although the spin texture becomes more complex than the unstrained case. This may be consistent with the quantitatively similar magnitude and symmetry of $\chi_{yxx}$ between films grown on GdScO$_3$ and NdGaO$_3$ despite the large difference in strain.

Finally, we remark on the contribution from disorder. In Fig. S9, we show the temperature dependence of the magnitude of the nonlinear planar Hall effect for the SrIrO$_3$ film grown on LSAT (001) substrate. The nonlinear planar Hall signal increases with decreasing temperature, exhibiting nearly the same trend as mobility shown in Fig S9(c), which would indicate disorder origin. However, recent theoretical studies showed that, unlike the spin Hall and anomalous Hall effects in the linear-response regime, the nonlinear Hall effect follows the dependence linear in scattering time $\tau$ both in the intrinsic and the extrinsic cases, which makes it difficult to differentiate these contributions [39,40]. As the nonlinear planar Hall effect [17] has to do with the nonlinear Hall effect observed at zero magnetic field [14,15] in terms of Berry phase dipole in the intrinsic case, we expect that both phenomena may also be similarly influenced by the disorder. As suggested in Ref. [40], precise disorder tuning would be necessary to separate these contributions.

In conclusion, we have observed the nonlinear planar Hall effect in nominally centrosymmetric SrIrO$_3$ thin films at room temperature, which is expected only in noncentrosymmetric systems with nontrivial spin textures in the momentum space. Comparing the nonlinear planar Hall signal between SrIrO$_3$ films grown on different substrates, we conclude that the observation originates from complex the spin texture in the two Dirac bands triggered by the antisymmetric spin-orbit interaction. The strength, sign, and anisotropy of the nonlinear planar Hall signal are modulated by the in-plane symmetry of the substrates, which may sensitively reflect anisotropy of the spin texture of the Dirac bands. Our study suggests that the nonlinear planar Hall effect can be observed in more diverse materials with strong spin-orbit interaction by engineering spin texture in the momentum space with utilizing epitaxial strain in the reduced dimensions. A similar concept has been theoretically proposed for the nonlinear spin Hall effect in strained graphene and Dirac semimetal [13,41]. We expect future theoretical studies will clarify how to effectively design symmetry breaking to induce a large nonlinear spin-charge conversion.

We acknowledge A. Kurita for the technical assistance of thin film growth, Y. Toyooka for the synthesis of the target, and J. Uzuhashi for transmission electron microscopy. This work was supported by Grants-in-Aid for Scientific Research (B) No. 19H02604. This work was partly carried out at NIMS Nanofabrication Platform and NIMS Namiki Foundry

# Supplemental Material for "Observation of Large Nonlinear Spin-Charge Conversion in the Thin Film of Nominally Centrosymmetric Dirac Semimetal SrIrO$_3$ at Room Temperature"

## EXPERIMENTAL DETAILS

The SrIrO$_3$ thin films are grown at 650 °C under an oxygen partial pressure of 100 mTorr by pulsed laser deposition using the 4th harmonic of an Nd:YAG laser (LS-2145TF, LOTIS TII) with a repetition rate of 5 Hz and an energy fluence of 15 mJ. The crystal structures of SrIrO$_3$ thin films are confirmed by X-ray diffraction.

For electrical transport measurement, a Hall bar structure is defined by photolithography and ion milling followed by the deposition of Ti/Au electrodes, which results in a typical channel width of 10 μm. The second-order term of the planar Hall effect is characterized by an out-of-phase second-harmonic component using lock-in amplifiers with an alternating current frequency of 33 Hz. The signal is found to be almost independent of the excitation frequency up to around 1 kHz, indicating that parasitic capacitance is not significant.

## RECIPROCAL SPACE MAPPING

The in-plane relationship of SrIrO$_3$ thin films and substrates is characterized by reciprocal space mapping around (103) diffraction of a pseudocubic representation as shown in Fig. S1. The SrIrO$_3$ thin film grown on GdScO$_3$ substrate is fully strained, whereas the films grown on LSAT and NdGaO$_3$ are overall strained but partially relaxed. The low intensity of the SrIrO$_3$ film grown on NdGaO$_3$ is due to the low crystal quality as also seen in the $\theta$-$2\theta$ scan in Fig. 1(b).

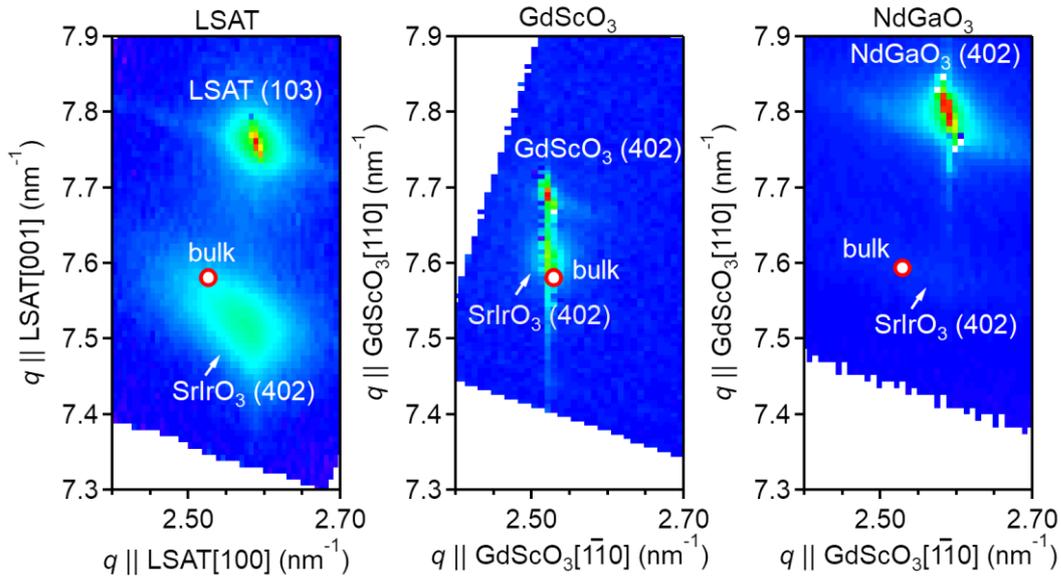

FIG. S1: Reciprocal space mapping of the SrIrO$_3$ thin films grown on LSAT (001), GdScO$_3$ (110), and NdGaO$_3$ (110) substrates. The reciprocal space point of the bulk SrIrO$_3$ is indicated by a red circle.

## SCANNING TRANSMISSION ELECTRON MICROSCOPY

In addition to X-ray diffraction, the crystal structures of SrIrO$_3$ thin films grown on LSAT (001), GdScO$_3$ (110), and NdGaO$_3$ (110) substrates are observed by the high-angle annular dark-field scanning transmission electron microscope (HAADF-STEM) images as shown in Fig. S2. In the case of GdScO$_3$ (110) substrate, SrIrO$_3$ thin film is found to show coherent lattice matching without noticeable dislocation, while dislocations and twins are partly observed for LSAT (001) and NdGaO$_3$ (110) substrates. LSAT has a cubic structure unlike orthorhombic SrIrO$_3$ and therefore twin is inherently expected in the film. For NdGaO$_3$, the crystal is isostructural with SrIrO$_3$ but a relatively large lattice mismatch may cause the dislocations at the interface as seen in Fig. S2(c).

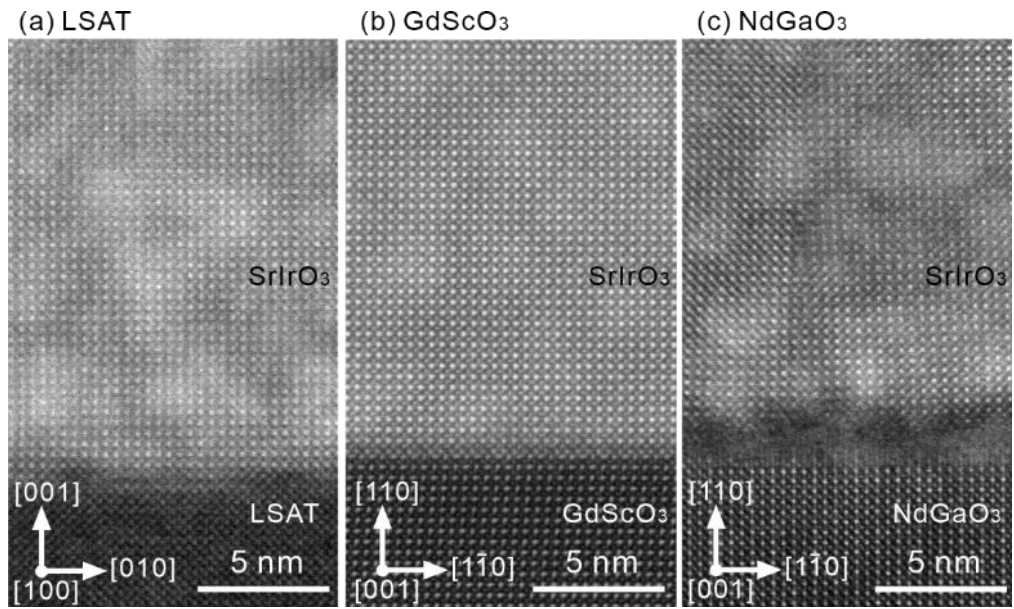

FIG. S2: High-angle annular dark-field scanning transmission electron microscope images for SrIrO$_3$ thin films grown on (a) LSAT (001), (b) GdScO$_3$ (110), and (c) NdGaO$_3$ (110) substrates.

**CURRENT AND MAGNETIC FIELD DEPENDENCES OF THE NONLINEAR PLANAR HALL EFFECT**

Here we show the current and magnetic field dependences of the nonlinear planar Hall effect for SrIrO3 thin films grown on LSAT (001) ($I \parallel [100]$) and GdScO3 (110) ($I \parallel [001]$). The data in Fig. S3 clearly show that the magnitude of $\Delta R_{yx}^{2\omega}$ is increases with both current and magnetic field. $\Delta R_{yx}^{2\omega}$ for the GdScO3 substrate is plotted as a function of current and magnetic field in Figs. S3(e) and S2(f), respectively, indicating almost linear dependence as an indication of bilinear nature. For the case of GdScO3 substrate with a current direction $I \parallel [001]$, $\Delta R_{yx}^{2\omega}$ tends to saturate at a large current above ~ 2 mA. This indicates that another mechanism is present to restrict the nonlinear effect at a large current, but the reason is not clear at the moment.

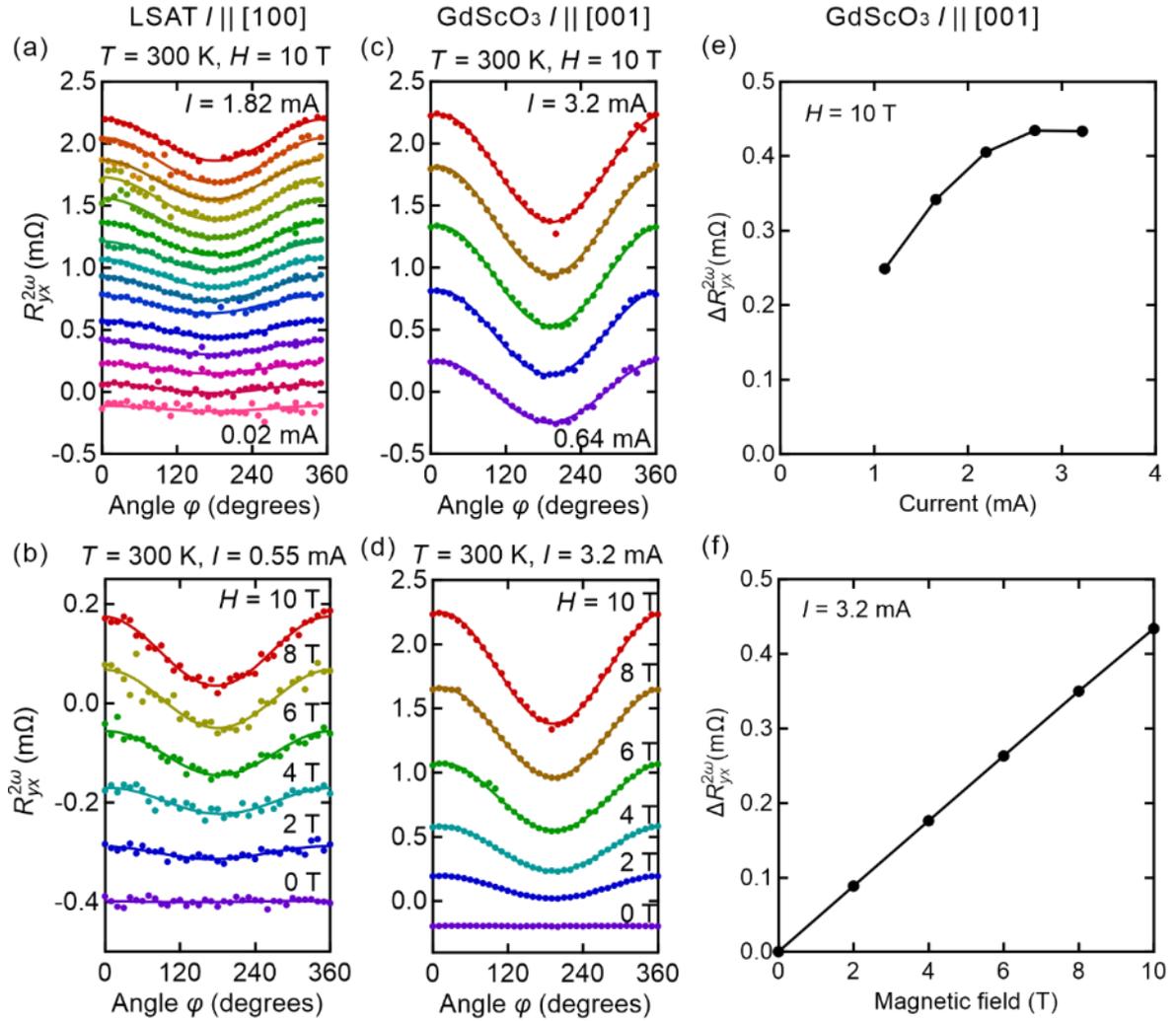

FIG. S3. (a) Current and (b) magnetic field dependences of raw data of $R_{yx}^{2\omega}$ for the film grown on LSAT substrate with a current direction along [100], and (c,d) those for the film grown on GdScO$_3$ substrate with a current direction along [001]. $\Delta R_{yx}^{2\omega}$, estimated from the magnitude of the cos$\varphi$ fitting for the data in (c) and (d), is plotted as a function of (e) current and (f) magnetic field for GdScO$_3$. The same plot for LSAT substrate is shown in Figs. 2(c) and 2(d) in the main text. The data in (a)–(d) are shifted vertically for clarity.

## BILINEAR MAGNETOELECTRIC EFFECT IN THE LONGITUDINAL RESISTANCE

The second-harmonic component of the anisotropic magnetoresistance ($R_{xx}^{2\omega}$) is shown in Fig. S4 together with the nonlinear planar Hall effect ($R_{yx}^{2\omega}$) for the SrIrO$_3$ thin film grown on LSAT (001) substrate, measured at 100 K under 8 T. Figure S4 shows the sin$\varphi$ dependence of $R_{xx}^{2\omega}$ and cos$\varphi$ dependence of $R_{yx}^{2\omega}$ as in the case for Bi$_2$Se$_3$ in Ref. [17]. In general for the SrIrO$_3$ films, $R_{xx}^{2\omega}$ is much noisier than $R_{yx}^{2\omega}$ probably due to larger background first-harmonic component, and $R_{xx}^{2\omega}$ is not discernable unlike $R_{yx}^{2\omega}$ at room temperature. The ratio $\rho_{yx}^{2\omega}/\rho_{xx}^{2\omega} = \left(R_{yx}^{2\omega}/R_{xx}^{2\omega}\right)(L/W) = 0.50$ (*L*: channel length = 25 μm, *W*: channel width = 10 μm) indicates that the second-harmonic signals do not originate from the Nernst effect, which is expected to give $\rho_{yx}^{2\omega}/\rho_{xx}^{2\omega} = 1$ [17].

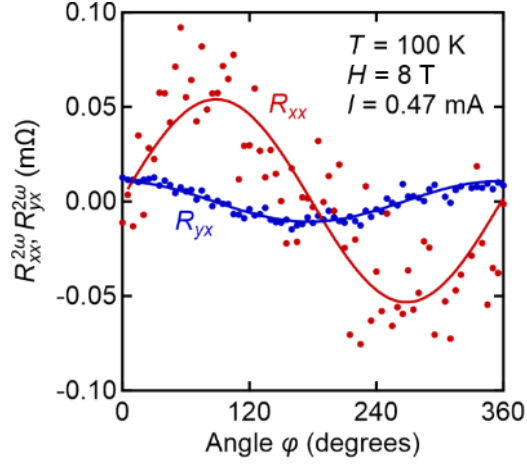

FIG. S4. Second-harmonic anisotropic magnetoresistance ($R_{xx}^{2\omega}$) and planar Hall effect ($R_{yx}^{2\omega}$) for the SrIrO$_3$ film grown on LSAT (001) substrate, measured at 100 K under 8 T with a current of 0.47 mA.

**THICKNESS DEPENDENCE OF THE NONLINEAR PLANAR HALL EFFECT**

We have investigated thickness dependence of the nonlinear planar effect for SrIrO$_3$ films grown on LSAT (001) substrate as shown in Fig. S5. To obtain an insight into the contribution to the nonlinear planar Hall signal from the surface layer, two kinds of normalization are compared by using the two-dimensional unit, $R_{yx}^{2\omega}$, and the three-dimensional unit, $\rho_{yx}^{2\omega} = R_{yx}^{2\omega} d$, where $d$ is the thickness of the film. The three-dimensional unit is appropriate when the bulk contribution is dominant, while the two-dimensional unit assumes an interface or surface origin. Figures S5(c) and S5(d) indicate that $\Delta R_{yx}^{2\omega}$ normalized by electric field $E_x$ and magnetic field $H$ is more or less independent of thickness compared with $\Delta\rho_{yx}^{2\omega}$, which is suggestive of the dominant contribution of interface or surface in our samples.

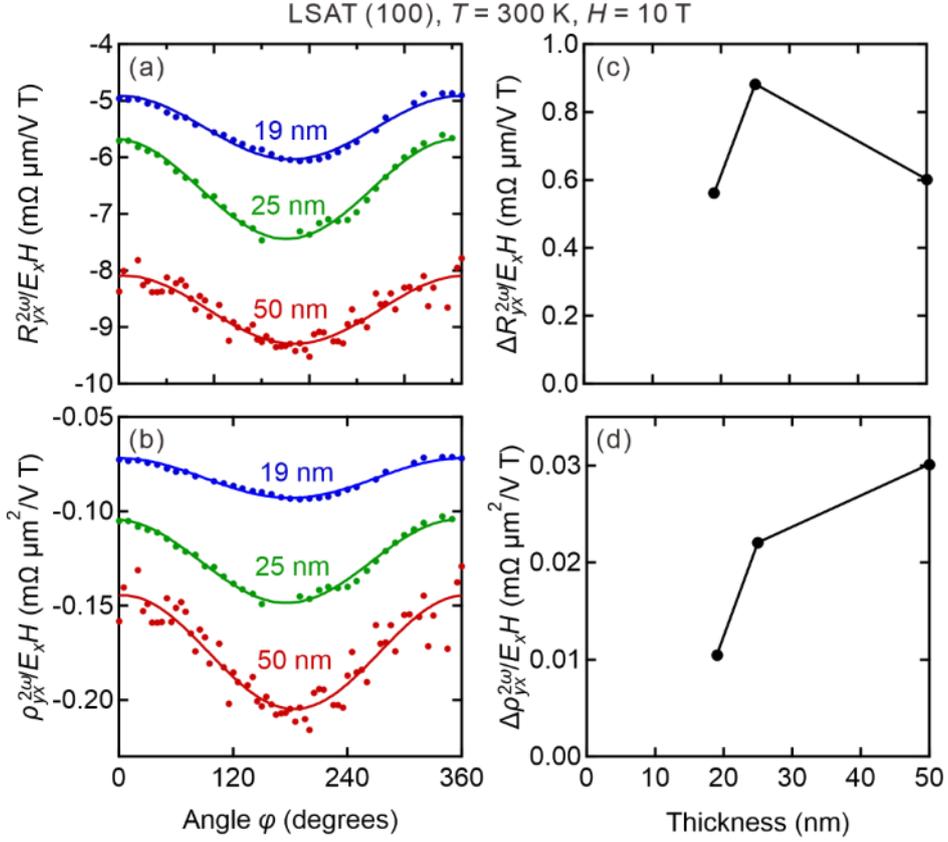

FIG. S5. (a,b) Nonlinear planar Hall measurement for $SrIrO_3$ thin films grown on LSAT (001) substrate with three different thicknesses. (c,d) Thickness dependence of the amplitude of nonlinear planar Hall signal normalized by electric field $E_x$ and magnetic field $H$. Two kinds of plots (a,c) in the two-dimensional unit and (b,d) in the three-dimensional unit are shown for comparison.

## NONLINEAR PLANAR HALL MEASUREMENT FOR A $SrRuO_3$ FILM

As a control experiment, we have performed the nonlinear planar Hall measurement for a SrRuO$_3$ thin film (thickness: 13 nm) grown on an LSAT (001) substrate. SrRuO$_3$ is a typical paramagnetic metal among oxide materials at room temperature (in contrast to the low-temperature ferromagnetic phase). As shown in Fig. S6, the cos$\varphi$ dependence is not discernable for the SrRuO$_3$ thin film compared with SrIrO$_3$ thin film, measured under similar conditions ($T$ = 300 K and $H$ = 8 T). Although the signal of SrRuO$_3$ is noisier due to higher resistance, the absence of the nonlinear planar Hall signal is clear.

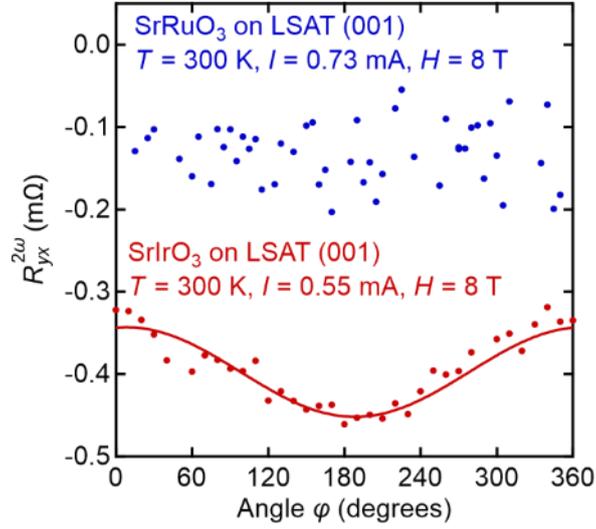

FIG. S6: Comparison of the nonlinear planar Hall measurements for SrIrO$_3$ and SrRuO$_3$ thin films grown on LSAT (100) substrate, measured at 300 K under 8 T. The data are shifted vertically for clarity.

**FIRST-PRINCIPLES BAND STRUCTURE CALCULATION**

The first-principles density-functional calculations including the spin-orbit interaction are carried out with the aid of the Vienna ab initio simulation program (VASP) [29]. Here, the generalized gradient approximation (GGA) [30] is adopted for the exchange-correlation energy and the projector augmented wave (PAW) potential [31] is used to treat the effect of core electrons properly. A plane-wave cut-off energy of 500 eV is employed for the wave function and $8 \times 8 \times 2$ $k$ points ($11 \times 11 \times 8$ $k$ points for bulk) are used for the Brillouin-zone integration. In this study, on-site Coulomb interaction $U$ is not included as $U$ is known not to qualitatively change the result [19].

To understand the observed nonlinear planar Hall Effect based on spin texture in the momentum space, we have carried out the first-principles band calculations for bulk [Fig. S7(a)], and unstrained [Fig. S7(b)] and strained [Fig. S7(c)] SrIrO$_3$ thin films. For the unstrained and strained cases, a slab model consisting of four SrO-IrO$_2$ stacks with SrO terminations is used as more stacks need unrealistically long calculation time. To construct the slab model, the in-plane orthorhombic $[1\bar{1}0]$ direction is redefined with a lattice constant of $a_s$ (7.75 Å for the unstrained slab), while [001] direction is kept the same as bulk. For the strained film, we have defined the slab unit cell based on the crystal structure of the SrIrO$_3$ thin film grown on GdScO$_3$ (110), where the detailed crystal parameters are taken from Ref. [38], which is monoclinic with lattice parameters of $a$ = 5.6120 Å, $b$ = 5.5865 Å, $c$ = 7.934 Å, and $\gamma$ = 90.367° (angle between $a$ and $b$ axes) and redefined lattice parameters of $a_s$ = 7.93 Å. The corresponding Brillouin zone is shown in Fig. S7(d).

The band structures of the bulk and the slab models are shown in Figs. S7(e)–S7(j)

along the red lines shown in Fig. S7(d). As a result of the redefinition of the unit cell for the slabs, we take the equivalent cut in the Brillouin zone shown in Fig. S7(d) because the band dispersion along the [110] direction is projected to (110) plane. For the slabs, contributions from the surface (top and bottom) Ir layers and inner Ir layers are represented by red and blue colors, respectively. The Fermi surfaces are found to be composed mainly of the surface (top and bottom) Ir layers. The conduction bands of bulk seem to be split into several copies of bands due to the different local environment of Ir ions.

The spin texture originating from the top Ir layer and the bottom Ir layer for the slab models are separately drawn around $\bar{U}$ in Fig. S8. There are four Fermi surfaces and all eigenvalues at each $k$ point are doubly degenerate due to the degeneracy of top and bottom surfaces. To compare the calculation with the experiment, contributions from Ir ions are projected to top and bottom Ir layers separately as shown in Fig S8, where the arrows indicate local spin direction at each $k$ point. Here the length of the arrows is arbitrarily scaled to show the relative in-plane component. We find spin textures are present in $k$ space in the slab model, while the spin component exactly cancels out by integration over the Fermi surfaces, reflecting the nonmagnetic nature. Also, the spin directions are exactly inverted between the contributions from top and bottom Ir layers, reminiscent of antisymmetric spin-orbit interaction. The strained case is also understood qualitatively in a similar way, but the Fermi surfaces are enlarged compared with the unstrained case probably due to the larger bandwidth originating from stretched Ir-O-Ir bonds, which enhances the transfer integral. Accordingly, spin texture is also changed in a complex manner, but the qualitative structure is similar particularly near the $\overline{\text{U}}$ point.

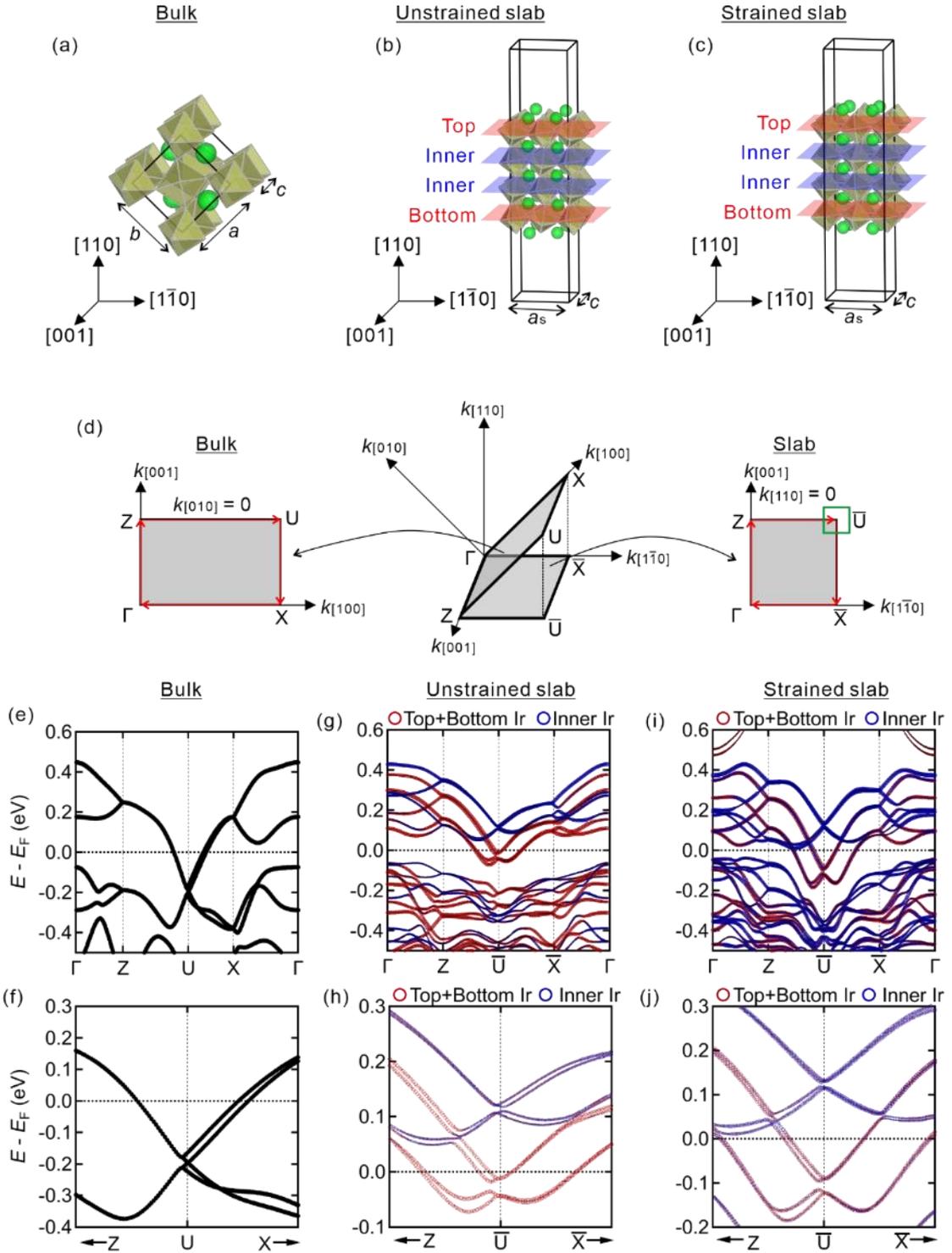

FIG. S7: Schematic diagrams of crystal structures for (a) bulk, (b) unstrained slab, and (c) strained slab. Black frames indicate the unit cells, including vacuum layers for the slab models. (d) Correspondence between Brillouin zones for the bulk and slab model. Due to the redefinition of the unit cell in the slab, the corresponding cut line in the Brillouin zone for the slab is a projection to (110) plane from the bulk band. The green square indicates the area of the Brillouin zone shown in Fig. S8. Band structures of (e,f) bulk, (g,h) unstrained slab, and (i,j) strained slab. The band structures around the $\bar{U}$ point are magnified in (f), (h), and (j). For the slab band structures, contributions from the surface (top and bottom) Ir layers and inner Ir layers are classified by color.

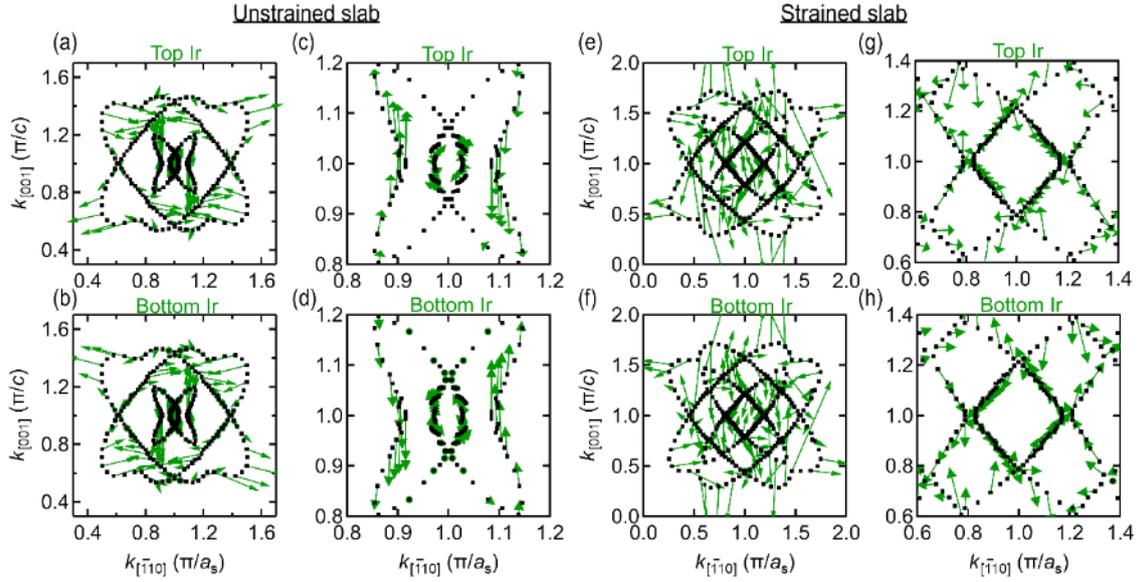

FIG. S8: Fermi surfaces around the $\bar{U}$ point [green square shown in Fig. S7(d)] for (a–d) unstrained, and (e–h) strained slab models. Local spin structures are shown by separately projecting the contributions from top Ir and bottom Ir layers. The Fermi surfaces are magnified around $\bar{U}$ in (c), (d), (g), and (h).

## TEMPERATURE DEPENDENCE OF CARRIER DENSITY, MOBILITY AND

# NONLINEAR PLANAR HALL EFFECT

The temperature dependence of the nonlinear planar Hall effect, carrier density, and mobility is measured as shown in Fig. S9. The nonlinear planar Hall signal shows nearly the same temperature dependence as the mobility. This tendency is qualitatively consistent with a theoretical prediction of $\tau$-linear dependence of $\Delta R_{yx}^{2\omega}$, where $\tau$ is the carrier scattering time.

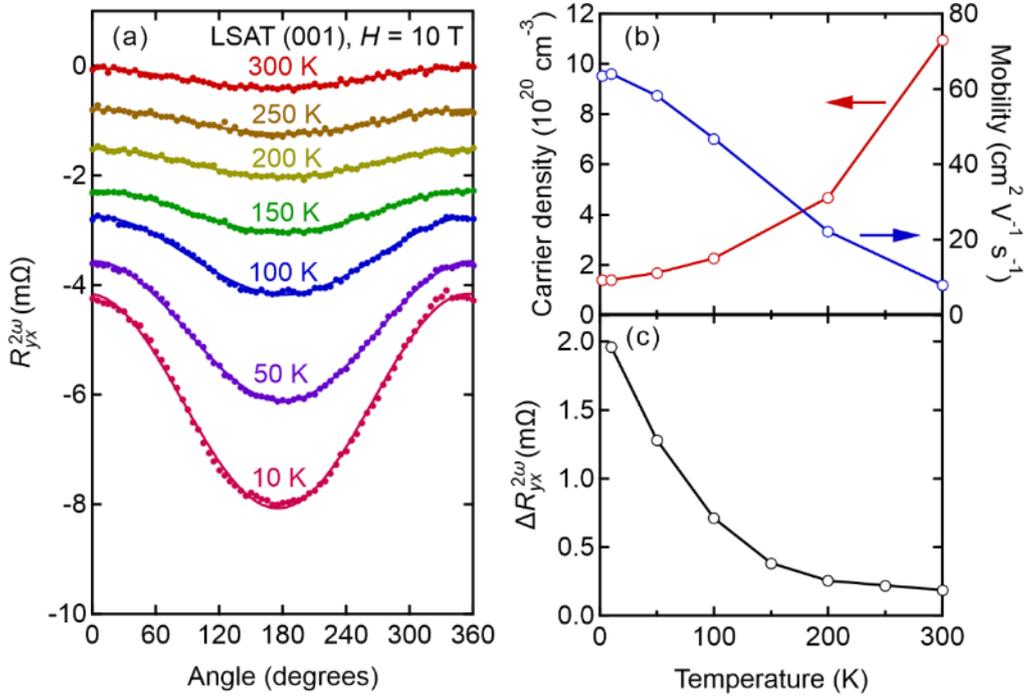

FIG. S9. (a) Temperature dependence of the nonlinear planar Hall effect measured at $H = 10$ T for the SrIrO$_3$ film grown on LSAT (001) substrate. The data are shifted vertically for clarity. Temperature dependence of (b) carrier density, mobility, and (c) magnitude of the nonlinear planar Hall effect.